\documentclass[12pt,preprint]{aastex}
\usepackage{graphicx}

\newcommand\lsim{\lower0.5ex\hbox{$\; \buildrel < \over \sim \;$}}
\newcommand\gsim{\lower0.5ex\hbox{$\; \buildrel > \over \sim \;$}}
\newcommand{\nus}{\mbox{$\nu_{\rm s}$}}
\newcommand{\nul}{\mbox{$\nu_1$}}
\newcommand{\nuu}{\mbox{$\nu_2$}}
\newcommand{\dnu}{\mbox{$\Delta\nu$}}
\newcommand{\rin}{\mbox{$R_{\rm in}$}}
\newcommand{\ro}{\mbox{$R_0$}}
\newcommand{\win}{\mbox{$\omega_{\rm in}$}}
\newcommand{\ws}{\mbox{$\omega$}}
\newcommand{\bz}{\mbox{$B_{\rm z}$}}
\newcommand{\bfi}{\mbox{$B_{\phi}$}}

\shorttitle{A Model for Twin kHz QPOs in NS LMXBs}
\shortauthors{Li \& Zhang}

\begin{document}
\title
{A MODEL FOR TWIN KILOHERTZ QUASI-PERIODIC OSCILLATIONS IN NEUTRON
STAR LOW-MASS X-RAY BINARIES}

\author
{X.-D. Li\altaffilmark{1} and C.-M. Zhang\altaffilmark{2}}

\affil{$^1$Department of Astronomy, Nanjing University,
    Nanjing 210093, China; lixd@nju.edu.cn}
\affil{$^2$National Astronomical Observatories, Chinese Academy of
Sciences, Beijing 100012, China; zhangcm@bao.ac.cn}

\date{}


\begin{abstract}
We suggest a  plausible interpretation for the twin kiloHertz
quasi-periodic oscillations (kHz QPOs) in neutron star low-mass
X-ray binaries. We identify the upper kHz QPO frequencies to be
the rotational frequency and the lower kHz QPOs the standing kink
modes of loop oscillations at the inner edge of the accretion
disk, respectively. Taking into account the interaction between
the neutron star magnetic field and the disk, this model naturally
relates the twin QPO frequencies with the star's spin frequencies.
We have applied the model to four X-ray sources with kHz QPOs
detected simultaneously and known spin frequencies.

\end{abstract}


\section{INTRODUCTION}
The kiloHertz quasi-periodic oscillations (kHz QPOs) have been
measured in more than twenty neutron star low-mass X-ray binaries
(NS LMXBs) in their persistent emission with the {\em Rossi X-ray
Timing Explorer} ({\em RXTE}),  which  offered unique insights
into the physics of strong gravity and dense matter (see van der
Klis 2000, 2005 for reviews). In many cases the twin kHz QPOs
appear simultaneously and the correlations between the pair
frequencies have been investigated extensively (e.g., Psaltis et
al. 1998, 1999b; Belloni et al. 2005).
Moreover, their frequencies also follow rather tight correlations
with other timing features of the X-ray emission (Ford \& van der
Klis 1998; Psaltis et al. 1999a; Belloni et al. 2002).
There is currently no consensus as to the origin of these QPOs,
nor on what physical parameters determine their frequencies, which
have been identified with various characteristic frequencies in
the inner accretion flow (see e.g. Stella \& Vietri 1999;
Osherovich \& Titarchuk 1999; Lamb \& Miller 2001; Abramowicz et
al. 2003).

With the discovery of the twin kHz QPOs in the  accretion-powered
millisecond pulsar SAX J1808.4$-$3658, it was found that the
frequency separation $\dnu$ is almost half the spin frequency
(Wijnands et al. 2003). For other sources with detected spin
frequencies $\nus$ from the burst oscillations, $\dnu$ are shown
to be close to either the spin frequencies or half of them (van
der Klis 2005 and references therein).
  %
%
These findings seem to hint some underlying mechanisms relating
$\nus$ to  the upper and lower kHz QPO frequencies ($\nuu$ and
$\nul$). However, the  more detailed measurements show that $\dnu$
is generally inconsistent with a constant value of $\nus$, but
varying with $\nuu$ or $\nul$ (van der Klis 2000, 2005 and
references therein), which cast doubts about the validity of the
simple beat-frequency interpretation (Miller, Lamb \& Psaltis
1998; see Lamb \& Miller 2001 for a modified version).  Osherovich
\& Titarchuk (1999) suggested the lower kHz QPO frequency $\nul$
to be the Keplerian frequency in the disk and the higher kHz QPO
frequency $\nuu$ the hybrid between $\nul$ and $2\nus$. Klu\'zniak
et al. (2004) showed that the twin kHz QPOs can be explained by a
nonlinear resonance in the epicyclic motion in the accretion
disks, which can lead to the 3:2 ratio for the two main resonances
(see also Abramowicz \& Klu\'zniak 2001; Abramowicz et al. 2003).


In this paper we  propose  an alternative interpretation for the
origin of the twin kHz QPOs, by considering the interaction
between the neutron star magnetic field and the surrounding
accretion disk. We introduce the model in \S 2, and present its
applications to several NS LMXBs with the simultaneously detected
kHz QPOs and known spin frequencies in \S 3. The possible physical
implications and conclusions are given in \S 4.

\section{MODEL}
Neutron stars in LMXBs generally accretes via an accretion disk.
For most part of the disk, the plasma rotates in a Keplerian
orbit. Close to the NS surface, the stellar magnetic fields begin
to truncate the disk and control the motion of the plasma,
resulting in a non-Keplerian boundary layer lying between the
magnetosphere corotating with the star and the outer Keplerian
disk. We assume that the boundary layer is confined by the inner
and outer radii, $\rin$ and $\ro$ respectively. As conceivable,
the plasma corotates with the magnetosphere at $\rin$, and the
plasma's motion begins to deviate from Keplerian rotation and take
its maximum value at $\ro$ (see Fig.~1).

As for the construction of the model, {\em we identify the upper
kHz QPO frequency $\nuu$ to be the rotational frequency at $\ro$},
i.e.,
\begin{equation}
\nuu\equiv\nu(\ro)\equiv\xi\nu_{\rm K}(\ro),
\end{equation}
where $\nu_{\rm K}(\ro)$ is the Keplerian rotation frequency at
$\ro$, and $0<\xi\lsim 1$. The value of $\xi$ depends on the
rotational frequency distribution inside the boundary layer.
Unfortunately, there is no analytic solution to the structure of
the boundary layer, and the value of $\xi$ can be evaluated only
through numerical calculations.
However, based on the qualitative characteristics of the rotation
rate in the boundary layer, Campbell (1987) has suggested the
following form for the disk rotation profile close to the
magnetosphere
\begin{equation}
\nu(R)=\nu_{\rm K}(R)-\nu_{\rm K}(\rin)(1-\win)
\exp[\frac{3(R/\rin-1)}{2(1-\win)}],
\label{nur}
\end{equation}
where $\win=\nus/\nu_{\rm K}(\rin)$.
When $R\gg\rin$,  $\nu(R)\rightarrow\nu_{\rm K}(R)$; when
$R\rightarrow\rin$, $\nu(R)\rightarrow\nus$.
Equation (2) gives a reasonable description of $\nu(R)$ close to
$\rin$, but our interest here focuses on the rotational behavior
around $\ro$, at which  Eq.~(2) fails to be valid for a
sufficiently wide range of $\win$.
Instead, we take  a slightly modified version of Eq.~(2) to
account for disk rotation,
\begin{equation}
\nu(R)=\nu_{\rm K}(R)-\nu_{\rm K}(\rin)(1-\win)
\exp[\frac{2(R/\rin-1)}{(1-\win)}].
\label{nur2}
\end{equation}
As an illustration, Fig.~1 shows $\nu(R)/\nu_{\rm K}(\rin)$
against $R/\rin$ for three selected values of $\win=0.2$, 0.5 and
0.8 in solid curves. The dashed curve corresponds to Keplerian
rotation for comparison.
Although the hypothesized form (3) is phenomenological, it shares
the main features with the numerically calculated results (e.g.,
Ghosh \& Lamb 1979), and the following analysis also indicates
that its adequate for the disk rotation.

From Eq.~(3) we can determine the location of $\ro$ using the
condition ${\rm d}\nu/{\rm d}R=0$ at $\ro$, and obtain the values
of $\xi\equiv\nu(\ro)/\nu_{\rm K}(\ro)$ for different values of
the so-called fastness parameter $\ws\equiv\nus/\nu_{\rm K}(\ro)$.
Figure 2 shows $\xi$ as a function of $\ws$.
%

Before discussing  the origin of the lower kHz QPOs, we note that
in solar physics  the QPOs of several minute periods in coronal
loops have been detected and successfully interpreted as the
standing kink modes of MHD waves (Aschwanden et al. 1999;
Nakariakov et al. 1999).  Thus, following  the similar ideas in
treating  the QPOs in the coronal loops,  we suppose that these
MHD waves could also be excited at the inner edge of the accretion
disks, where the poloidal field lines are dragged along the
azimuthal direction due to shear motion between the star and the
disk, and various types of MHD instabilities take place (Ghosh \&
Lamb 1979). The reconnection of the azimuthal magnetic field lines
could result in circular flux tubes (loops) in the boundary layer,
where plasma is confined along a magnetic field line with some
cross section. It is well known  that the coronal loops may be set
into oscillations with various modes leading to brightness
oscillations (Roberts 2000 for a review). While the similar
oscillations may also occur in the accretion disks, generated by
MHD turbulence, here we focus on the fast kink mode of standing
MHD waves. These fast waves arise as a free mode only for high
density loops (Edwin \& Roberts 1982), which may be appropriate in
accretion disks in NS LMXBs. Assume the loop length being the
circumference of the magnetosphere, $2\pi\ro$, the oscillation
frequency is
\begin{equation}
\nu=\frac{Nc_{\rm k}}{4\pi\ro},
\end{equation}
where $c_{\rm k}$ is the kink speed. The wave number
$k=N\pi/(2\pi\ro)=N/(2\ro)$. The integer $(N-1)$ stands for the
node number of the vibration along the tube axis,
 with $N=1$ being the principal mode. At the inner edge of
the accretion disk, the  sound speeds are much smaller than the
azimuthal Alfv\'en speeds $c_{\rm A\phi}$, and  the plasma inside
a loop is much denser than its surroundings for the comparable
magnetic field strengths, therefore we have the kink speed
 $c_{\rm k}\simeq \sqrt{2}c_{\rm A\phi}$ (Roberts 2000),
and in turn the frequency of a standing
kink mode
\begin{equation}
\nu=\frac{Nc_{\rm A\phi}}{2\sqrt{2}\pi\ro}.
\end{equation}

In the following, we estimate the azimuthal field strength at
$\ro$. We assume that the stellar magnetic field lines are
initially dipolar and penetrate the accretion disk. The
differential rotation between the star and the disk generates the
azimuthal field component $\bfi$ from the vertical component $\bz$
(Ghosh \& Lamb 1979).
According to Wang (1995), if the growth of $\bfi$ is limited by
the diffusive decay produced by the turbulent mixing within the
disk, $\bfi$ is given by\footnote{Other processes related to the
dissipation of the magnetic field give similar expressions of
$\bfi$ (Wang 1995).}
\begin{equation}
\bfi(R)=\gamma\frac{\nu(R)-\nus}{\nu(R)}\bz(R),
\end{equation}
where the parameter $\gamma\sim 1$ (Aly 1984; Uzdensky,
K$\ddot{o}$nigl, \& Litwin 2002).

We assume that the magnitude of $\ro$ is close to that of the
Alf\'ven radius where the magnetic energy density equals the total
kinetic energy density (Davidson \& Ostriker 1973), i.e.,
\begin{equation}
\frac{\bz(R_0)^2}{8\pi}=\frac{1}{2}\eta\rho v_{\rm K}^2|_{R_0},
\end{equation}
where $\rho$ is the mass density, and $\eta\sim 1$.
Combining Eqs.~(6) and (7) we get the azimuthal Alf\'ven speed at
$\ro$,
\begin{equation}
c_{\rm A\phi}(\ro)=\gamma\eta^{1/2}\frac{(\nuu-\nus)}{\nuu}v_{\rm
K}(\ro).
\end{equation}


Furthermore, {\em we suggest the lower kHz QPOs to be the
principal fast kink mode of the standing MHD waves along the
$\bfi$ field lines at $\ro$}. Inserting Eq.~(8) into Eq.~(5) we
find the frequency to be
\begin{equation}
\nul=\gamma(\frac{\eta}{2})^{1/2}\frac{(\nuu-\nus)}{\nuu}\nu_{\rm
K}(\ro).
\end{equation}
Combining Eqs.~(1) and (9) we have
\begin{equation}
\nul=(\frac{\alpha}{\xi})(\nuu-\nus),
\end{equation}
where $\alpha=\gamma(\eta/2)^{1/2}\sim 1$ is taken as a free
parameter that absorbs the uncertainties in determining $\ro$ and
$\bfi(\ro)$. Since the $\xi(\ws)$ relation is known from Fig.~2,
Eq.~(10) suggests a unique relation between $\nul$ and $\nuu$ for
given values of $\nus$.

Obviously,  the MHD oscillations represent the  continuum, i.e.,
the oscillation frequency is a continuous function of a wave
number. We ascribe the lower kHz QPOs to be the principal fast
kink mode with the loop length of $L=2\pi\ro$ due to the following
facts. Firstly, the global modes with $N=1$ or 2 are easiest to
excite in the loops (Roberts, Edwin, \& Benz 1984). Secondly, for
a standing wave the decay time of the MHD oscillations is
$\tau\propto (L/N)^2$ (Roberts 2000). Apparently, the oscillations
with the longer $L$ and smaller $N$ last the longer lifetime, and
hence are more easily to be detected. Thus, for the fast kink mode
at $\ro$, oscillations along $\bfi$ have the maximum loop length
and smallest wave number when $L=2\pi\ro$ and $N=1$.
There may exist the other oscillation modes (e.g. the fast sausage
mode of standing fast MHD waves, standing slow MHD waves, and
propagating MHD waves) in the accretion disks, and some of them
could be also shown  as some kinds of QPOs. However, their
oscillation frequencies are either too high or too low compared to
those of the kHz QPOs in NS LMXBs. Additionally it seems that the
kink mode is most likely to be seen as a standing wave (Roberts
2000) .

\section{COMPARISON WITH OBSERVATIONS}
We have compared the predicted $\nul-\nuu$ relation with the
observations of kHz QPOs in four NS LMXBs (4U1608$-$52,
4U1636$-$53, 4U1728$-$34, and 4U1915$-$05), in which   both the
spin and twin kHz QPO frequencies have been measured (data were
provided by T. Belloni, M. M\'endez and D. Psaltis). In Fig.~3 the
crosses represent the measurements and the solid lines stand for
the theoretical relations. The spin frequency and the adopted
value of $\alpha$ for each source are also displayed. In all cases
the value of $\alpha$ is of order of unity as expected.
The theoretical predictions and the measured kHz QPO data  match
quite well for   4U 1728$-$34 and 4U 1915$-$05, albeit the
approximated consistence is obtained for 4U 1608$-$52 and 4U
1636$-$53, which may be caused by the fact that the real magnetic
field structure in these sources may be more complicated than in
the simplified model considered here.

\section{DISCUSSION AND CONCLUSIONS}
In this paper we have presented a plausible mechanism for the
production of twin kHz QPOs in NS LMXBs, invoking the magnetic
field-accretion disk interaction. Although NSs in LMXBs are
thought to have weak magnetic fields ($\sim10^{8-9}$ G), their
influence on the disk rotation has not been paid much attention in
the existed works on this subject. Titarchuk, Lapdius, \& Muslimov
(1998) has already suggested that a shock occurs in the transition
layer where the Keplerian disk adjust to sub-Keplerian flow. The
disk can undergo various types of oscillations under the influence
of the gas, radiation, magnetic pressure and gravitational force.
As to the origin of the oscillations, the boundary layer in our
work is similar to but closer to the NS compared with the
centrifugal barrier region in Titarchuk et al. (1998).

One of the twin kHz QPO frequencies is usually interpreted as the
Keplerian rotation frequency at some preferred radii, most likely
the inner edge of the disk (e.g., Miller at al. 1998; Titarchuk \&
Osherovich 1999). We have followed this idea, but suggested that
the real (non-Keplerian) rotation at $\ro$ leads to the upper kHz
QPOs. Unfortunately, since there is no general analytic form for
the rotational profile within the boundary layer, we have to use a
phenomenological description for the rotation of disk plasma
around the inner edge\footnote{Even if we take $\xi=1$, that is,
Keplerian rotation at $\ro$, Eq.~(10) still fits the observational
data fairly well.}. Although it seems to be adequate, its validity
and accuracy should be testified more carefully in the future by
both observational and theoretical investigations.

Moreover, we interpret the lower kHz QPOs to be the fast kink
modes of MHD oscillations in loops along the $\bfi$ field lines at
$\ro$, shared with the similar physical mechanism for coronal loop
oscillations. Since $\bfi$ is generated by $\bz$ through shear
motion between the NS and the disk, this naturally links the twin
kHz QPOs with the stellar spin. But it is distinct from the
traditional beat-frequency model by the fact that in our work the
twin OPOs originate from different physical processes. We note
that similar hypothesis was discussed by Muslimov (1995), who
suggested that the QPOs in LMXBs may be caused by the excitation
of the so-called nonlinear global Alfv\'en modes in the boundary
layer plasma. It is interesting that these modes were observed in
the studies of an ideal MHD spectrum of a toroidal plasma
(Goedbloed 1975). In addition, the detailed numerical
investigation of these modes was performed by  Appert et al.
(1982), and their existence was confirmed experimentally by Behn
 et al. (1984) and Evans et al. (1984).

Recent analysis by Barret, Olive, \& Miller (2005) showed that the
quality factor for the lower kHz QPOs in 4U 1636$-$53 increases
with frequency up to a factor of $\simeq 200$ when $\nul\simeq
850$ Hz, then drops at higher frequencies. A ceiling of the lower
kHz QPO frequency at 920 Hz is also seen. In the frame of the
present work, we ascribe these features to the evolution of the
twisted field lines. Several theoretical studies of the star-disk
interaction (e.g. Aly 1984, 1985; Uzdensky et al. 2002) have shown
that as a dipole field is twisted due to differential rotation,
the field lines inflate and effectively open up when a critical
twist angle is attained (Uzdensky et al. 2002). This limits the
azimuthal pitch at $\ro$,
$|\bfi(\ro)/\bz(\ro)|=\gamma(\nuu-\nus)/\nuu$, to some critical
value, say, $\gamma_{\rm c}$, demonstrating that a steady state of
configuration could be established only if the rotation shear
$\nuu-\nus$ is small enough. These arguments suggest that there
may exist a maximum value of $\nuu$, beyond which the $\bfi$ field
becomes unstable, resulting in decreasing quality factor of the
lower kHz QPOs at higher frequencies and a saturation frequency of
$\nul$ when most of the field lines become open.

We have presented a qualitative description of the kHz QPO
production mechanism and a crude quantitative expression of the
kHz QPO frequencies, to interpret the observed kHz QPO phenomena.
As a preliminary exploration, many physical details have not been
considered, such as in what condition and how much MHD wave energy
is produced to account for the observed Fourier power spectrum of
kHz QPOs. These should be investigated more carefully in the
future work.

\acknowledgements This work was supported by NSFC under grant
number 10025314 and MSTC under grant number NKBRSF G19990754.
We are grateful for  T. Belloni,  M. M\'endez and D. Psaltis for
providing the QPO data, and T. P. Li and P. F. Chen for helpful
discussions.
%
The authors express thanks to an anonymous referee for the
critical comments that greatly helped improve the manuscript.

\clearpage
\begin{figure}
\includegraphics{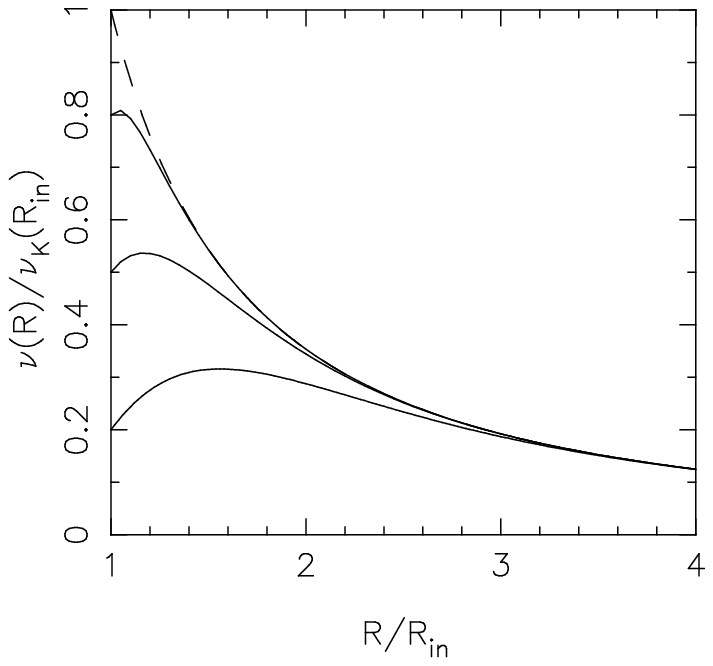}

\caption{The solid lines show the dimensionless rotational
frequencies in the accretion disk for $\win=0.2$, 0.5, and 0.8
(from bottom to top). The dashed line represents the Keplerian
rotation. }
\end{figure}

\clearpage
\begin{figure}
\includegraphics{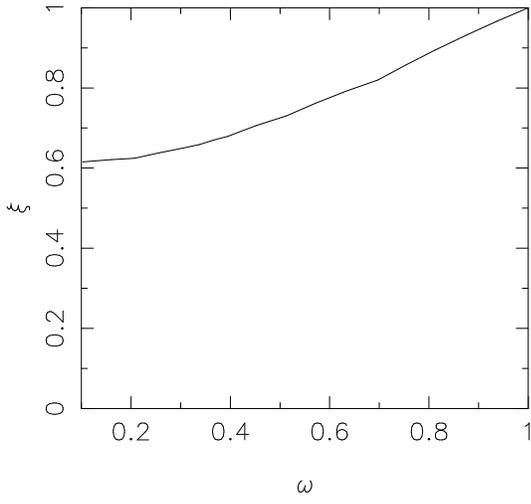}
\caption{The parameter $\xi$ plotted against the fastness
parameter $\ws$.}
\end{figure}

\begin{figure}
\includegraphics{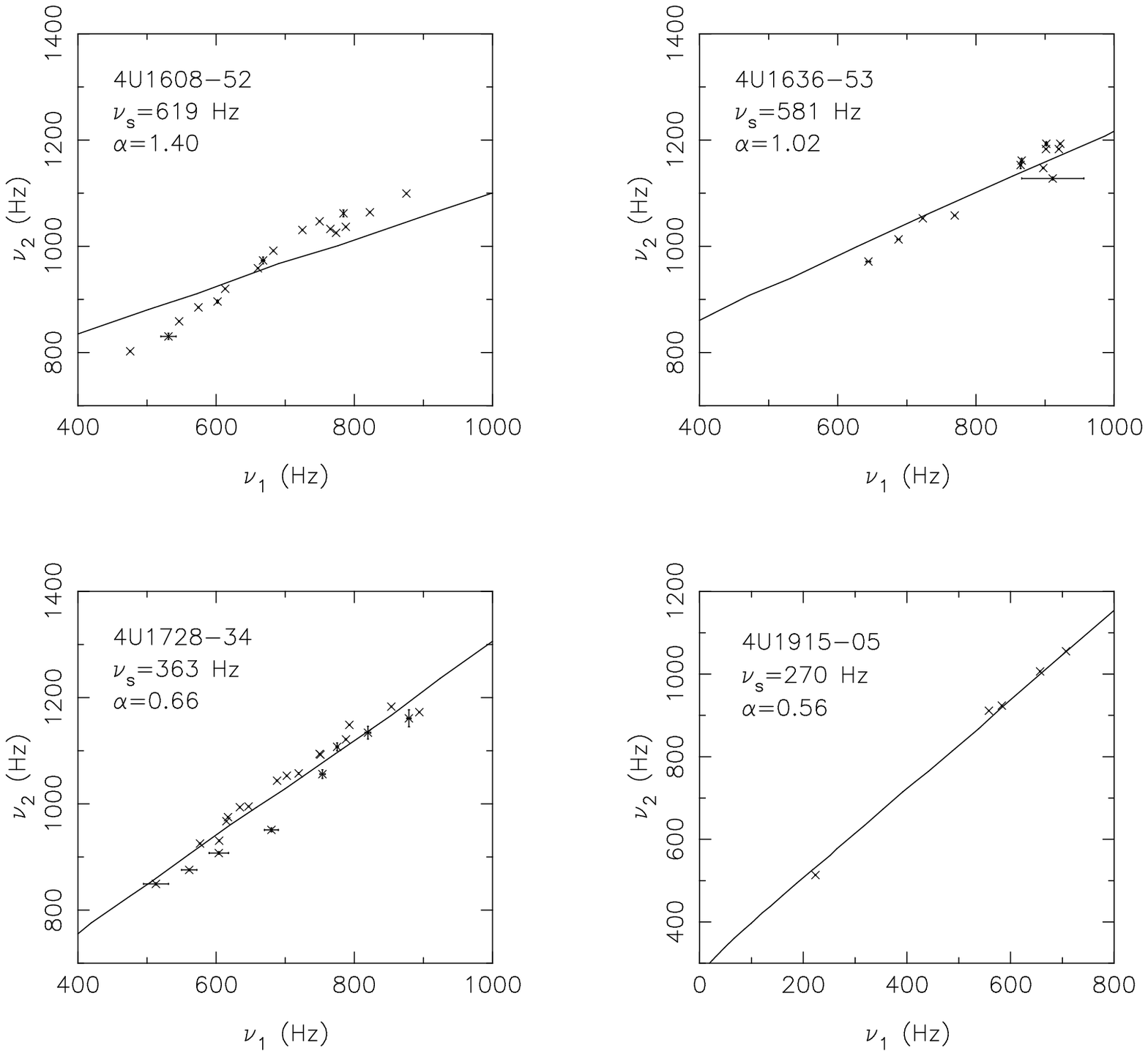}
\caption{Predicted relations for twin kHz QPO frequencies and
measured frequencies for four NS LMXBs}
\end{figure}


\begin{thebibliography}{}

\bibitem[2003]{ab03}
Abramowicz, M.~A., Bulik, T., Bursa, M., \& Klu\'zniak, W. 2003,
A\&A, 404, L21

\bibitem[2001]{ab01}
Abramowicz, M.~A. \& Klu\'zniak, W. 2001, A\&A, 374, L19

\bibitem[1984]{al84}
Aly, J. J. 1984, ApJ, 283, 349

\bibitem[1985]{al85}
Aly, J. J. 1985, A\&A, 143, 19

\bibitem[1982]{ap82}
Appert, K., Gruber, R., Troyon F., \& Vaclavik, J. 1982, Plasma
Phys., 24, 1147

\bibitem[1999]{as99}
Aschwanden, M. J. et al. 1999, \apj, 520, 880

\bibitem[2005]{ba05}
Barret, D., Olive, J.-F., \& Miller, M. C. 2005, MNRAS, 361, 855

\bibitem[1984]{be84}
Behn, R. et al.
1984, Plasma Phys. Controlled Fusion, 26, 173

\bibitem[2002]{be02}
Belloni, T., Psaltis, D.,  \& van der Klis, M., 2002, \apj, 572,
392.

\bibitem[2005]{be05}
Belloni, T.,  Mendez, M. \&  Homan, J.  2005, A\&A, 457, 209

\bibitem[1987]{ca87}
Campbell, C. G. 1987, MNRAS, 229, 405

\bibitem[1973]{da73}
Davidson, K. \& Ostriker, J. P. 1973, ApJ, 179, 585

\bibitem[1982]{ed92}
Edwin, P. M. \& Roberts, B. 1982, Sol. Phys. 76, 239

\bibitem[1984]{ev84}
Evans T. E. et al. 1984, in Proc. of the Fourth Joint
Varenna-Grenoble Internat. Symp. on Heating in Toroidal Plasma,
Rome, 1, 121

\bibitem[1998]{fo98}
Ford, E. C. \& van der Klis, M. 1998, ApJ, 506, L39

\bibitem[1979]{gh79}
Ghosh, P. \& Lamb, F. K. 1979, ApJ, 232, 259

\bibitem[1975]{go75}
Goedbloed, J. P. 1975, Phys. Fluids, 18, 1258


\bibitem[2004]{kl04}
Klu\'zniak, W., et al. 2004, ApJ, 603, L89



\bibitem[2001]{la01}
Lamb, F. K., \& Miller, M. C. 2001, \apj, 554, 1210



\bibitem[1998]{mi98}
Miller, M.\ C., Lamb, F.\ K., \& Psaltis, D.\ 1998, \apj, 508, 791


\bibitem[1995]{mu95}
Muslimov, A. G. 1995, in Physics of Neutron Stars, ed. A. M.
Kaminker, Nova Science Publishers, Inc.,   p. 277

\bibitem[1999]{na99}
Nakariakov, V. M. et al. 1999,  Science, 285, 862

\bibitem[1999]{os99}
Osherovich, V. \& Titarchuk, L.  1999,  ApJ, 522, L113


\bibitem[1998]{ps98}
Psaltis, D. et al.
1998, \apj, 501, L95

\bibitem[1999a]{ps99a}
Psaltis, D., Belloni, T. \& van der Klis, M. 1999a, ApJ, 520, 262

\bibitem[1999b]{ps99b}
Psaltis, D. et al.
1999b, \apj, 520, 763

\bibitem[2000]{ro00}
Roberts, B. 2000, Sol. Phys., 193, 139

\bibitem[2000]{ro84}
Roberts, B., Edwin, P. M,, \& Benz, A. O. 1984, \apj, 279, 857

\bibitem[1999]{st99}
Stella, L., \& Vietri, M., 1999,  \prl,  82, 17 


\bibitem[1998]{ti98}
Titarchuk, L., Lapidus, I., \&  Muslimov, A. 1998, \apj,  499, 315

\bibitem[1999]{ti99}
Titarchuk, L., \& Osherovich, V. 1999, ApJ, 518,



\bibitem[2002]{uz02}
Uzdensky, D. A., K$\ddot{o}$nigl, A., \& Litwin, C. 2002, ApJ,
565, 1191

\bibitem[2000]{vk00}
van der Klis, M. 2000, ARA\&A, 38, 717

\bibitem[2005]{vk05}
van der Klis, M. 2005, 
to appear in Compact stellar X-ray sources, eds.  W. H. G. Lewin
\& M. van der Klis, Cambridge University Press  (astro-ph/0410551)

\bibitem[1995]{wa95}
Wang, Y. M. 1995, ApJ, 449, L153

\bibitem[2003]{wj03}
Wijnands, R. et al.
2003, \nat,  424, 44

\end{thebibliography}
\end{document}